Karlsruhe Institute of Technology

**This is the author's version of a work that was published in the following source:**

Wagner, N., & Maedche, A. (2026). Designing For Imperfect Prompts: A Simulation-Based Evaluation of an LLM-Based Enterprise Data Assistant Using Knowledge Graphs. Proceedings of the 34th European Conference on Information Systems (ECIS 2026). 5.
https://aisel.aisnet.org/ecis2026/bus_analytics/bus_analytics/5

human-centered systems lab
https://h-lab.win.kit.edu/

Please note: Copyright is owned by the author and / or the publisher.
Commercial use is not allowed.

**© 2017. This manuscript version is made available under the CC-BY-NC-ND 4.0 license http://creativecommons.org/licenses/by-nc-nd/4.0/**

# DESIGNING FOR IMPERFECT PROMPTS: A SIMULATION-BASED EVALUATION OF AN LLM-BASED ENTERPRISE DATA ASSISTANT USING KNOWLEDGE GRAPHS

*Completed Research Paper*

Niklas Wagner, Karlsruhe Institute of Technology (KIT), Karlsruhe, Germany, niklas.wagner@kit.edu

Alexander Mädche, Karlsruhe Institute of Technology (KIT), Karlsruhe, Germany, alexander.maedche@kit.edu

## Abstract

*Enterprise data assistants using large language models (LLMs) promise democratized data access in organizations. Yet in our real-world deployment within the CFO unit's data lake at an industry partner, we observed that business users frequently pose imperfect prompts that include domain-specific* vocabulary *and lack schema references or structural information. This Design Science Research study examines automatic resolution using a semantic mediation layer. We design and evaluate two design variants: schema-based mediation via semantic similarity, and knowledge graph-based mediation leveraging an automatically constructed knowledge graph from the data lake. Our simulation-based evaluation using real-world data, supported by LLM-as-a-Judge assessments and human validation, shows that knowledge graph-based mediation outperforms the schema-based approach in Precision, Recall, F1, and preserving user intent. We propose a design principle for knowledge graph-based mediation, demonstrating that structured knowledge representation bridges user intent and database structure and guides the design of enterprise data assistants across skill levels.*

## 1 Introduction

Data-driven decision-making has become omnipresent and plays an increasingly important role in enterprises. This shift has created a growing demand for more intuitive and accessible information systems (IS) for users to interact with data. As SQL remains the dominant language for database systems, the interaction with data has evolved from structured querying of databases to natural-language interfaces powered by large language models (LLMs) (Liu & Xu, 2025). Introduced as data assistants, the most significant advancement of these systems lies in their ability to allow users to explore and analyse data without code writing. This introduces a new scientific consideration: as the retrieval of data becomes easier, user prompts directly determine the assistants' results. In practice, users may provide queries that seem clear and unambiguous to them, but from the perspective of the data assistant, they are incomplete, unstructured, or lack sufficient domain-specific knowledge (Yuan et al., 2023). We refer to such user prompts as imperfect prompts, as they often lead to conflicting, inaccurate, or inefficient data retrieval. As data assistants are becoming integral to data-driven decision-making, either users must learn how to effectively communicate with the system, or the system must proactively include a mediator that understands both the system's capabilities and the specific intent of its users.

In enterprise settings, business users rarely issue technically perfect prompts. Instead, prompts often describe analytical business goals rather than executable instructions, leading to misalignment between the user's actual intent and resulting database queries (Ning et al., 2024). Data sources often do not align with user expectations in terms of naming, scope, granularity, and terminology, reflecting the differences in vocabularies and conceptualizations across organizational subunits (Tasselli et al., 2020). LLMs attempt to mitigate linguistic ambiguities and bridge gaps in phrasing, but standalone, they cannot resolve misalignment between user intent, data semantics and domain-specific information (Wang et al., 2025). A lack of foundational knowledge can hinder effective interaction in multiple ways, including causing users to struggle with formulating appropriate prompts (Wissuchek & Zschech, 2025). Consequently, exposing data assistants to users who assume that knowledge is shared by the system and do not know how to pose their analytical business goals as structured prompts, increases the likelihood of inaccurate, incomplete, or inefficient data retrieval. In real-world enterprises, the skills and knowledge required by employees evolve continuously, often beyond what current training can fully address (Brasse et al., 2023). Hence, such systems demand a built-in mechanism to interpret, mediate, and account for imperfect prompts, going beyond what technical innovation alone can achieve. The idea of mediators in the context of IS is not new: A mediator is an intermediate layer that embodies encoded knowledge about data sources and uses that knowledge to simplify, abstract, merge, and explain data, making information more accessible (Wiederhold, 1992). Embedding models highlight the ability to retrieve relevant knowledge based on meaning rather than exact wording, adapting more to the user's vocabulary and intent. Knowledge graphs (KGs), in turn, capture and structure knowledge and their relationships, which is often necessary to interpret and determine data and their connections (Ji et al., 2022). As a result, leveraging KGs in a mediator is vital to transform data assistants from providing minor, isolated benefits into systems that deliver meaningful support to users across different skill levels.

Despite recent advancements in LLM-based enterprise data assistants, significant challenges remain unsolved. Research increasingly highlights the combination of KGs and embedding models for more nuanced knowledge retrieval, but much of this work focuses on knowledge bases themselves rather than on their integration within a mediator to interpret, contextualize, and resolve imperfect prompts of users. Evidence from text-to-SQL, KGs, and other domain-related research shows that incorporating explicit knowledge is not merely beneficial, but necessary for real-world enterprise scenarios, where data heterogeneity, ambiguity, and evolving user semantics cannot be resolved by LLMs alone (Huo et al., 2025). At the same time, enterprises increasingly employ structured and semi-structured platforms such as data lakes, with the goal to enable data to be accessed more transparently and through a variety of tools (Armbrust et al., 2021). KG-based mediators have been proposed in the literature for more efficient data and knowledge retrieval (Ohmori et al., 2024). Furthermore, KGs have also been proposed for schema matching and for direct mapping of relational databases to graphs (Ma et al., 2025; Boudaoud et al., 2023). However, less attention is given to the role of KGs in terms of the user and imperfect prompts. While KGs are increasingly applied to improve the retrieval of complex relational data, the need for mediation between source data and user intents persists. In contrast to the growing research interest in enterprise data assistants, few studies address the role of real imperfect user prompts and how they can be automatically enriched with a semantic mediation layer. To address this gap, we formulate our research question as follows:

*How can LLM-based enterprise data assistants be designed to automatically improve imperfect prompts of users using semantic mediation?*

To address this question, we conduct a Design Science Research (DSR) project following the framework of Peffers et al. (2007). Our DSR project is embedded in a real-world business environment within the CFO unit of our industry collaboration partner, a multinational energy corporation. In our prior work, we developed an LLM-based enterprise data assistant featuring dynamic few-shot examples, intent routing to specialized subagents, SQL-based database access, intermediate planning, and data visualization capabilities. Evidence from the first design cycle revealed persistent challenges related to users and their prompts: Prompts were short, never referenced database elements, and relied heavily on domain-specific vocabulary accumulated over years, terminology unfamiliar to the assistant. This issue persisted even though we provided simple explanations of the assistant's actions (including queries and database names) via an auxiliary LLM. In contrast, our evaluation showed that prompts explicitly referencing the correct schema elements produced reliable results. This finding motivates the focal point of our next design cycle. In the second cycle, we address the challenge of imperfect user prompts and their impact on interactions with the enterprise data assistant. To bridge user intent and source data, we introduce a semantic mediation unit as an intermediate layer and evaluate two architectural variants: (1) schema-based mediation, which retrieves relevant database schemas based on user prompts, and (2) KG-based mediation, which uses a KG automatically constructed from the database schema. We instantiate and assess both variants in a simulation-based evaluation using user interactions (*N=158*) from employees of our industry partner (*N=29*). Results show that KG-based semantic mediation outperformed the schema-based approach in terms of prompt enrichment and alignment with user intent. We evaluated the completeness of user intent using an LLM-as-a-Judge, and these assessments were validated by two human annotators from the industry partner. Quantitatively, the KG-based method achieved higher *Precision*, *Recall*, and *F1-Score* than the schema-based mediation. Additional analysis using mediated prompts with the main assistant from the first design cycle revealed nuanced insights. For prompts that already contained all relevant schema information, additional schema retrieval was obligatory, making data retrieval more efficient. For prompts missing some information, schema retrieval remained necessary and led to broader but correct results. Finally, prompts containing incorrect information led to erroneous outputs regardless of the mediation approach, underscoring the need for careful system design before deployment to end users.

The remainder of this paper is structured as follows. The next section presents the foundations and related work on data assistants, imperfect user prompts, mediating components in IS, and KGs. We then outline our DSR framework and describe the individual phases of our second design cycle. Afterward, we introduce our semantic mediation layer, which aims to bridge the gap between user intent and the underlying database structure. The evaluation draws on log data from our first design cycle and includes statistical analyses, an LLM-as-a-Judge, and assessments by two human annotators from our industry partner. Finally, based on the evaluation of the second design cycle, we present our resulting design principle, discuss the theoretical and practical implications of our work, and highlight limitations and directions for future research.

# 2 Foundations and Related Work

## 2.1 Conversational Data Assistants

Conversational data assistants are conversational agents that bridge natural-language intent and data retrieval or analysis tasks (Xu et al., 2025). They emerged from research on natural language interfaces to databases, which aim to allow users to query data without explicit knowledge of programming. Recent advances in LLMs have transformed these systems into more context-aware, multimodal assistants capable of natural language preprocessing and intent understanding, with capabilities to reason over table schemas and generate executable queries (Liu & Xu, 2025). However, real-world tasks such as data analysis demand domain-specific knowledge, including industry terminology, abbreviations, and jargon (Shi et al., 2025). Without this knowledge, LLMs often misinterpret queries or produce inaccurate responses. Domain adaptation can be achieved through techniques such as prompt engineering and fine-tuning, though both approaches face significant challenges. In addition, in-context learning is proposed to ground LLMs in enterprise information, improving accuracy and relevance (Gartner, 2023).

User prompts in data-oriented interactions are rarely syntactically wrong, yet they are often semantically or pragmatically incomplete (Kumar et al., 2024). Beyond ambiguity or missing schema mappings, imperfection stems from the gap between a user's analytical goal and its operational formulation (Kimiya & Huang, 2022). Furthermore, the demand for assistants to anticipate user needs and to provide meaningful, personalized interactions continues to grow (Kimiya & Huang, 2022). Many users express a clear analytical intent (e.g., "find top companies over time") but lack the vocabulary or schema-awareness needed to convert it into executable logic (Longxu et al., 2022). Moreover, users often rely on mental-model shortcuts: they simplify or compress instructions to match their internal representation of how the system works, which can lead to semantic ambiguity or system misunderstanding (Johnson-Laird, 1980; Popescu et al., 2003). Developments in research benchmarks illuminate these prompt imperfections: The Spider dataset provided a landmark cross-domain text-to-SQL benchmark but largely assumes grammatically clean, unambiguous user queries (Yu et al., 2018). Spider 2.0 extends the dataset realism to enterprise-scale databases and agents, including multi-query workflows, demanding searches through database metadata and large schemas, yet its prompts remain carefully curated rather than organically written (Lei et al., 2025). AmbiQT (Bhaskar et al., 2023) explicitly introduces ambiguity of user prompts by including examples where each utterance maps to SQL-statements due to lexical or structural ambiguity, forcing systems to reason over multiple interpretations. AMBROSIA (Saparina & Lapata, 2024) advances this further by cataloguing scope ambiguity, attachment ambiguity, and vagueness in real-style questions with human-annotated SQL interpretations. PRACTIQ (Dong et al., 2025) goes a step further by introducing conversational, real-world queries that are not only ambiguous but also sometimes unanswerable. Clarification is proposed as one way to resolve ambiguity but needs to consider the demands of the individual user and be based on given results (Zhang & Choi, 2025). BIRD-INTERACT shifts the focus by highlighting the role of external knowledge and metadata files, demanding that models do not solicit clarifications but instead account for external knowledge dynamically during interaction (Huo et al., 2025). Our first design cycle reflects these benchmark developments: Users included enterprise- and domain-specific jargon, lacked systematic structure, and used shorthand expressions rather than executable goals. Imperfection rarely signals multiple valid interpretations from users' point of view and is best addressed through knowledge-driven mediation.

## 2.2 Mediators and Knowledge Graphs

Mediators have long been recognized in IS as knowledge-specific modules that reconcile heterogeneous data sources. Wiederhold (1992) defines a mediator as a module that exploits encoded knowledge about certain sets or subsets of data to create information for a higher layer of applications. Mediators act as middleware, encoding domain knowledge to resolve ambiguities, overlapping scopes, and redundancy across sources (Wiederhold, 1992). This knowledge can be formalized in structured, maintainable forms, enabling the system to capture and reuse organizational knowledge (Holsapple & Joshi, 2004). Although mediation is often approached as a technical integration problem, traditional mediation focuses on aligning existing data sources, while semantic mediation extends this concept by explicitly representing the meaning (Karnstedt et al., 2003; Sonntag & Porta, 2014). Explicitly representing database semantics through structured knowledge models, such as KGs, allows semantic mediators to align user intentions with underlying data schemas (Putrama & Martinek, 2022). This is pivotal because user terminology often diverges from database schema terms, creating a "semantic gap" (Baik et al., 2019).

KGs are directed graphs composed of nodes representing entities (real-world objects or abstract concepts) and edges representing semantic relationships between them (Ji et al., 2022). KG embedding techniques transform entities and relations into low-dimensional vector spaces, preserving the structural and semantic information of the graph and enabling computational operations such as link prediction, node retrieval, or semantic matching (Kazemi & Poole, 2018; Gesese et al., 2021). In the context of relational databases, KGs can be leveraged in two primary ways: either to represent the data entities as a KG or to map semantics of a database (Boudaoud et al., 2023; Ma et al., 2025). The first approach may capture more detailed information but is complex and resource-intensive, whereas the second is simpler to implement but provides only a partial view of the underlying data and does not account for all relational constraints or integrity rules.

# 3 Research Method

For our research method, we adopt the DSR process model by Peffers et al. (2007) to guide our research project and its design process. Following a problem-centered initiation (Peffers et al., 2007), we conduct our research through consecutive design cycles that iteratively explore the design of our LLM-based enterprise data assistant. Figure 1 highlights these design cycles, illustrating how insights from each cycle inform the next. This paper encompasses the second design cycle of our DSR project. Our DSR project is grounded in the real-world, with the fundamental objective to empower business users in the CFO unit of our industry partner to access data and make informed decisions using natural language. In our previous work, we proposed a first version of the enterprise data assistant and piloted it with our industry partner. We designed an LLM-based system with secure SQL-based access to the unit's database and tool support for data exploration, including automated visualization and intermediate analytical planning steps. Guided by insights from the pilot, elicited user requirements, and related work, we adopted a modular multi-agent design in which a routing component assigns tasks to specialized subagents based on topic and estimated effort, reflecting the high heterogeneity of user requests in both subject matter and required analytical depth (Yue et al., 2025). Additionally, we found that dynamically retrieved few-shot examples from a domain-specific vector-store help condition the agents and improve response efficiency by reducing the need for extensive analytical reasoning by the agents (Izacard et al., 2023). This was necessary because, during the design and development stage, we found that users did not typically pose explicit analytical questions. Consequently, the core problem motivating this second design cycle was shaped by the interplay between users and the subsequent engineering effort required to translate user prompts into concrete, executable analytical tasks for our enterprise data assistant.

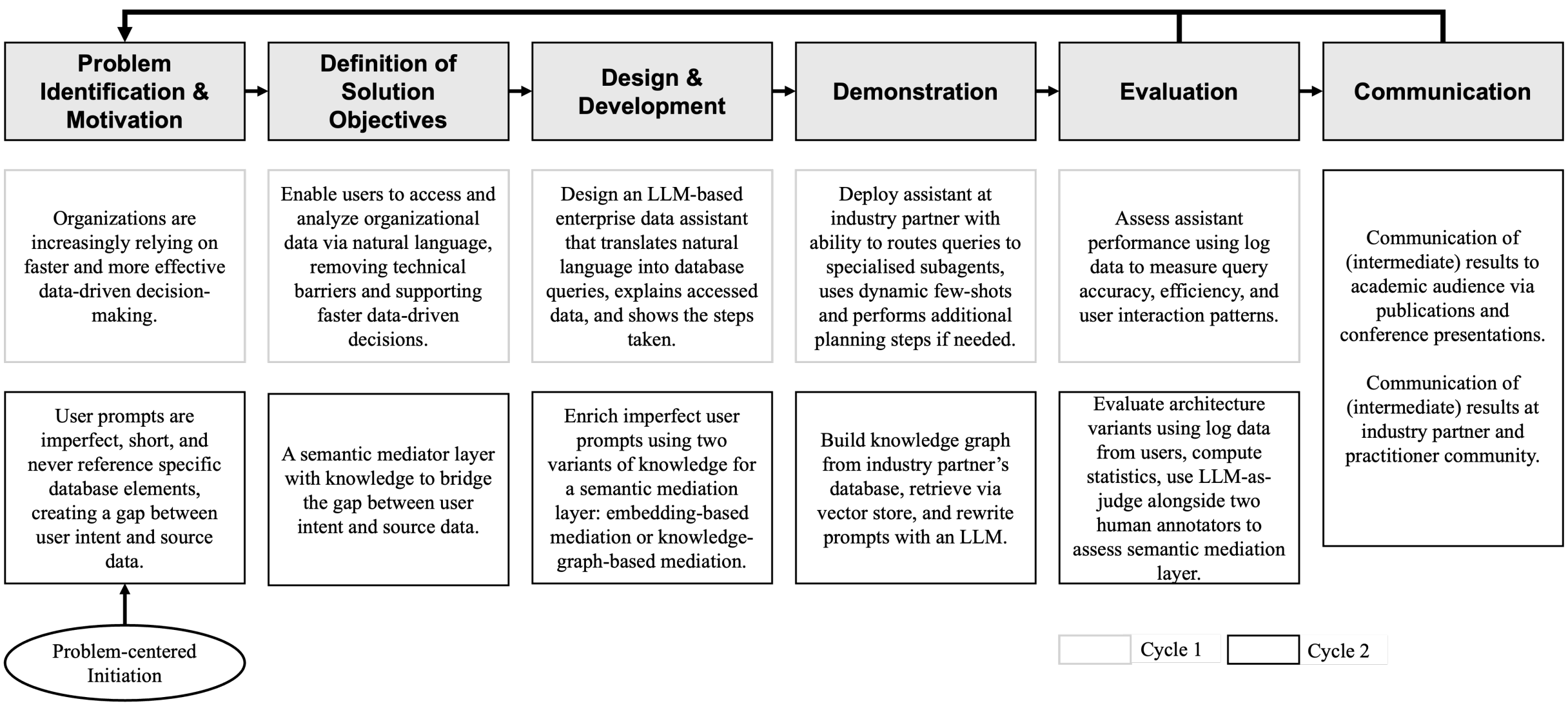

*Figure 1. DSR Framework.*

**Problem Identification and Motivation.** We observed multiple problems in how users interacted with our LLM-based enterprise data assistant and examined them through interaction log analysis and interviews. After an initial piloting period, we enabled detailed logging, added an auxiliary LLM to generate contextual hints, and translated tool outputs from raw SQL into natural-language explanations for users. Despite these changes, user prompts remained brief and seldom referenced concrete database schema elements, instead relied on company-specific terminology and wording conventions developed over years. Moreover, users reported that the assistant appeared to start each query "from scratch," performing full contextual searches without "leveraging previously gathered knowledge". This behaviour led to noticeable latency and gave users the impression of inefficiency, especially when users "repeated similar questions a few days later". To bridge this gap, we augmented the database elements with richer metadata, split our agent into three subagents based on task and effort, added an LLM-based router to delegate each request to the specific subagent, dynamically integrated few-shot examples using an additional vector-store, and invested in prompt-engineering strategies, which helped the overall system to align more quickly with user intent and the underlying database context. The LLM-based router directs each prompt to the subagent best suited for its domain, assigning simple queries to a fast subagent and complex queries to one of two more thorough subagent, each specialized in a different subdomain of the database. For example, queries that require quick retrieval of basic information are handled by the fast subagent, while queries demanding in-depth analysis are routed to a thorough subagent. Given controlled tests, we found that assistant performance improved markedly despite this engineering effort when user prompts explicitly referenced concrete schema elements, which motivated our next design cycle.

**Definition of Solution Objectives.** As a solution objective, we introduce a semantic mediation layer that grounds user input using database-native semantics. To enable this, we propose to automatically generate a lightweight KG from the database schema, capturing tables, columns, and relationships without any external domain knowledge or manual enrichment. KGs enable the representation of database knowledge in a connected form and facilitate operations such as relationship traversal and join-path inference over relational structures. Rather than modelling individual data entities, which would create a large, complex, and hard-to-maintain graph, our focus lies on schema-level elements. This approach keeps the KG compact and efficient while providing sufficient information for database-grounded prompts and remaining easy to update when the database changes.

**Design & Development.** For the design and development phase, we design an initial implementation of the semantic mediation layer and a corresponding schema-level KG that mitigates the need for dynamic few-shot examples through an additional vector-store (see Figure 2) as introduced in the first design cycle. For this, we design and compare two variants of our semantic mediation layer: First, a KG-based mediation, which uses the automatically generated KG from the database, and second, a schema-based mediation, which retrieves relevant database schemas based on semantic similarity. This design allows us to empirically assess whether the additional structure and reasoning capabilities of KGs provide a practical advantage over simpler schema-retrieval mechanisms in mediating user intent and database semantics, as our research is grounded in the real-world.

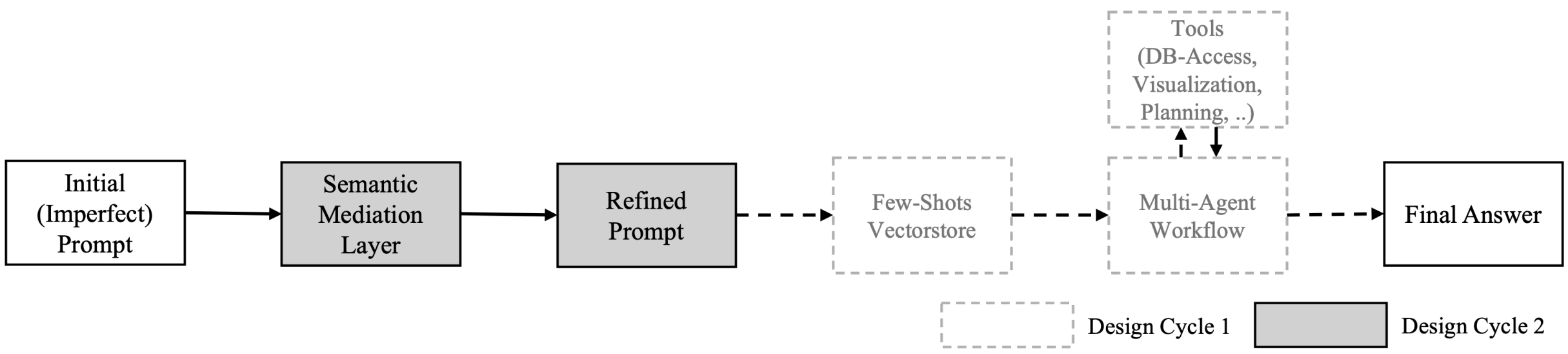

*Figure 2. Enterprise Data Assistant Architecture with Semantic Mediation Layer.*

**Demonstration.** In the first design cycle, we piloted our enterprise data assistant within the CFO unit of our industry partner. In this study, the focus shifts to evaluating the newly introduced mediator in a controlled simulation environment using user interactions derived from interaction logs within our first design cycle. The demonstration illustrates the concept of a semantic mediation layer, which decomposes user prompts into manageable chunks, identifies relevant knowledge pieces from the database context, and consolidates them into a coherent representation. In the KG-based mediation, relevant knowledge chunks are retrieved from our designed KG and expanded to capture the relationships present in the database, enabling an LLM to rewrite prompts with enriched domain knowledge. This process demonstrates how grounded mediation can improve alignment between user intent and database semantics without requiring manual intervention.

**Evaluation.** We base our analysis on interaction log files from 29 users and 158 recorded interactions captured during the most recent 30 days of system use. Although the assistant has been available for six months, only the last 30 days of logs are stored and therefore form the basis of our evaluation of the two design approaches. We analyse the extent and richness of the mediated prompts by computing word counts and employ LLM-based methods to assess the completeness of captured user intent. Importantly, we measure performance against a baseline to confirm that user intent is preserved even when database information is explicitly exposed. To ensure reliability, two human annotators evaluated the mediated outputs in two rounds of assessment. Additionally, two different LLMs are used in the evaluation process to compare the performance and consistency of our semantic mediation layer across the KG-based and schema-retrieval-based variants.

Based on the results of the simulation study and evaluation, we conclude the second design cycle of this study by synthesizing key insights regarding the effectiveness of semantic mediation for the imperfection of user prompts. The findings indicate that KGs provide richer structural context and alignment with user intent compared to schema-based mediation, while remaining efficient and maintainable. From these insights, we formulate our design principle of a knowledge-based semantic mediation layer for enterprise data assistants.

# 4 Problem Identification and Solution Objectives

In the upcoming section, we will address the issues identified in the piloting of our enterprise data assistant in our first design cycle and align them with findings from literature to derive concrete solution objectives for a semantic mediation layer. We first articulate the types of prompt imperfections observed with our artifact and their root causes in enterprise settings. We then translate the observations into requirements that guide our design.

## 4.1 Problem Identification

Our log analysis and user interviews revealed a systematic mismatch between how users expressed their analytical goals and the information our enterprise data assistant required to produce correct, efficient queries. User prompts were extremely brief, on average, 1.14 sentences and roughly 10 words, and did not reference schema elements such as table or column names, keys, or KPIs. Because the prompts stated high-level goals rather than executable instructions, the assistant could not reliably infer which entities, measures, and relationships were needed, and it frequently guessed or defaulted to superficially plausible but incorrect columns and joins. A recurrent failure pattern was the assistant’s tendency to anchor on similarly named attributes instead of recovering the structural logic implied by the user’s intent. A prominent example concerned the concept of VAT fiscal unity, where the correct interpretation requires establishing which subsidiaries are controlled by which parent entities based on shareholding and control relationships. In our data, a table contained a VAT fiscal unity state attribute, and in the absence of explicit guidance the assistant incorrectly used this column as a proxy for the concept, ignoring the necessary discovery of subsidiaries and shareholders and the associated join paths to hierarchy or ownership tables. In other words, the assistant selected the right-sounding column rather than reconstructing the relational structure that operationalizes the VAT group concept. Compounding this, the term fiscal unity is polysemous in our enterprise context: our data lake contains both income-tax and VAT fiscal unities. While domain experts readily disambiguate the intended meaning based on the organizational subunit or the involved entity type, the assistant lacked explicit cues.

To mitigate these issues in the first design cycle, we invested substantial engineering effort. We introduced a router that dispatched each query to one of three specialized subagents and augmented the system with a vector-store containing at least two few-shot examples per table. This configuration helped in domains we explicitly modeled, improving alignment where examples closely matched user wording. However, the approach remained brittle and labor-intensive: only previously modeled cases benefited, and questions that had not been anticipated still yielded incorrect answers. This reliance on extensive hand-curation is unsustainable at enterprise scale and undermines organizational effectiveness, highlighting the need for a principled, maintainable mechanism that grounds underspecified prompts in database-native semantics and resolves imperfections without exhaustive manual engineering. Across the 158 interactions from recent log data, none contained explicit schema references, and all user prompts were short and single-intent, underscoring the need for a method that bridges natural-language intent and database structure while remaining generalizable.

## 4.2 Solution Objective

To address the semantic gap between users' intent and how source data is structured, we propose to add a semantic mediation layer to our enterprise data assistant. The objective is to transform brief, goal-oriented prompts into database-grounded inputs without burdening users with schema knowledge or increasing operational overhead. The mediator must act as a controlled augmentation mechanism: it preserves the user's original intent while adding just enough database semantics to make execution of the assistant efficient, accurate, and reliable. From our analysis of the enterprise context, we derive three primary requirements:

**Preserve user intent**. Maintain the original goal, scope, and constraints of the user's prompt, ensuring that no critical meaning is altered or lost.

**Knowledge grounded**. Enrich prompts with the necessary domain and data semantics to resolve prompt imperfection and support mapping to the underlying data.

**Lightweight design**. Enable automatic, low-maintenance operation that integrates seamlessly with the existing enterprise data assistant.

We address these requirements by using a KG derived from the database schema as a source of structured knowledge. This graph captures the key elements of the data environment that are relevant to interpreting user intent, including the tables and their columns, descriptions or metadata that provide semantic context, and the relationships that connect tables, such as foreign key links. Relevant knowledge from this graph could be incorporated into the assistant to support reasoning over the user's request which enables a KG-based mediation variant, and to leverage structured knowledge to interpret and link domain knowledge dynamically.

# 5 Design & Development

Following the objective outlined in the previous section, we designed a semantic mediation layer that maps user intents to relevant database elements. In this design cycle, we focus on two complementary mediation variants, each illustrating a distinct approach to incorporating knowledge from a database.

**KG-based Mediation.** This variant leverages a KG automatically constructed from the database schema, encoding tables, columns, and their interrelationships. This allows the representation of direct and indirect joins, and more accurate retrieval of relevant nodes.

**Schema-based Mediation.** This variant adopts a simpler approach, directly retrieving relevant table schemas based on semantic similarity. While it does not exploit the graph structure or reasoning capabilities of KGs, it provides a lightweight and efficient baseline for comparison, highlighting the practical benefit of additional semantic structure.

## 5.1 Designing a Semantic Mediation Layer

Building on our insights into user intent and database structure from our industry partner, we observed that the objectives behind user prompts frequently involve combining data from multiple tables and understanding their relationships. In addition, user prompts typically describe the higher-level analytical goal rather than specifying schema or table names. To address this, we propose the design of a semantic mediation layer that operates through a multi-step process. First, the user prompt is segmented into distinct units, each corresponding to a specific aspect of the user's intent. For each segment, relevant knowledge is retrieved from either the underlying schema or a KG. In the KG-based variant, this retrieval is further enhanced by traversing the graph to identify connected tables and columns, capturing both direct and indirect relationships. This ensures a comprehensive understanding of the structural context and dependencies inherent in the data. Finally, an LLM is employed to refine the prompt, integrating the retrieved knowledge and structural insights into a final prompt that is both aligned with the user's intent and consistent with the source data. Figure 3 illustrates the overall design and workflow of our semantic mediation layer.

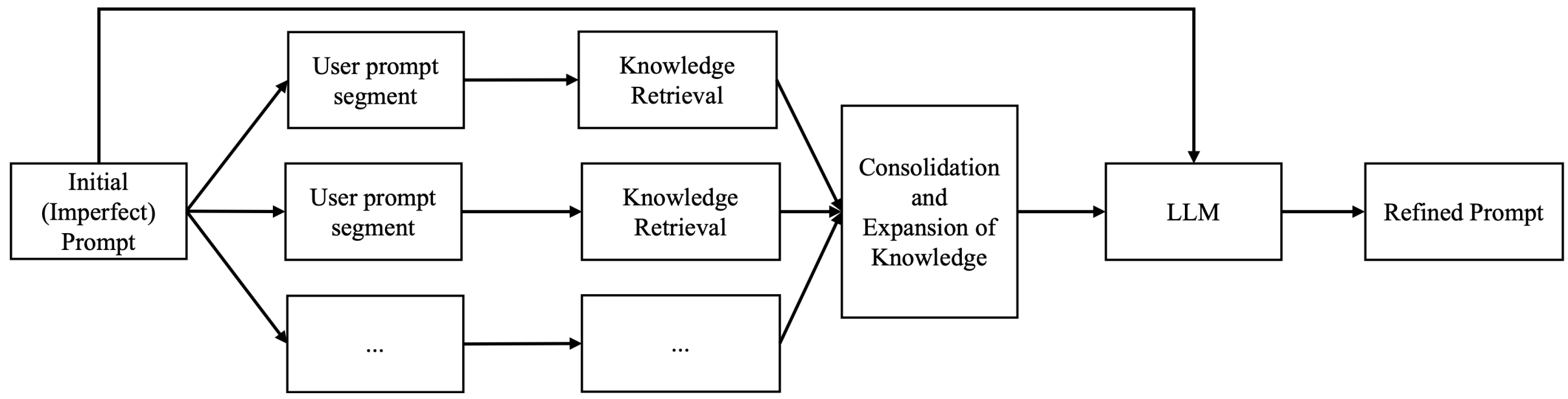

*Figure 3. Design of the Semantic Mediation Layer.*

## 5.2 Knowledge Graph Conceptualization

The conceptualization of the KG forms the foundation of our mediation approach and is detailed below. For the KG-based mediation approach, we automatically construct a lightweight KG from the underlying database schema, encoding the structural elements necessary to interpret user intent and link it to the corresponding data sources using the NetworkX 3.5 package. The KG contains two primary node types: Tables, representing the core entities in the database, and columns, representing the attributes within each table. Both node types are extracted directly from the schema. Relationships between these elements are modeled through *has_column* edges, which link tables to their constituent columns, and *joins_on* edges, which capture joinable relationships inferred from primary key–foreign key constraints or, when such constraints are absent, from heuristics such as matching names or compatible data types.

Although the KG is automatically generated and does not include domain-specific constructs such as specialized metrics or fine-grained filters, this design choice is deliberate: the goal is to isolate the contribution of structural database knowledge, rather than to achieve full domain completeness. The resulting graph provides sufficient relational structure to support reasoning over tables and columns while remaining lightweight, consisting of 273 nodes and 540 edges using the database of our industry partner. The required schema metadata was retrieved through Databricks Information Schema. Figure 4 illustrates the KG concept, showing how the graph structure enables the retrieval of direct neighbors, the detection of joinable columns, and the traversal of indirect multi-hop join paths.

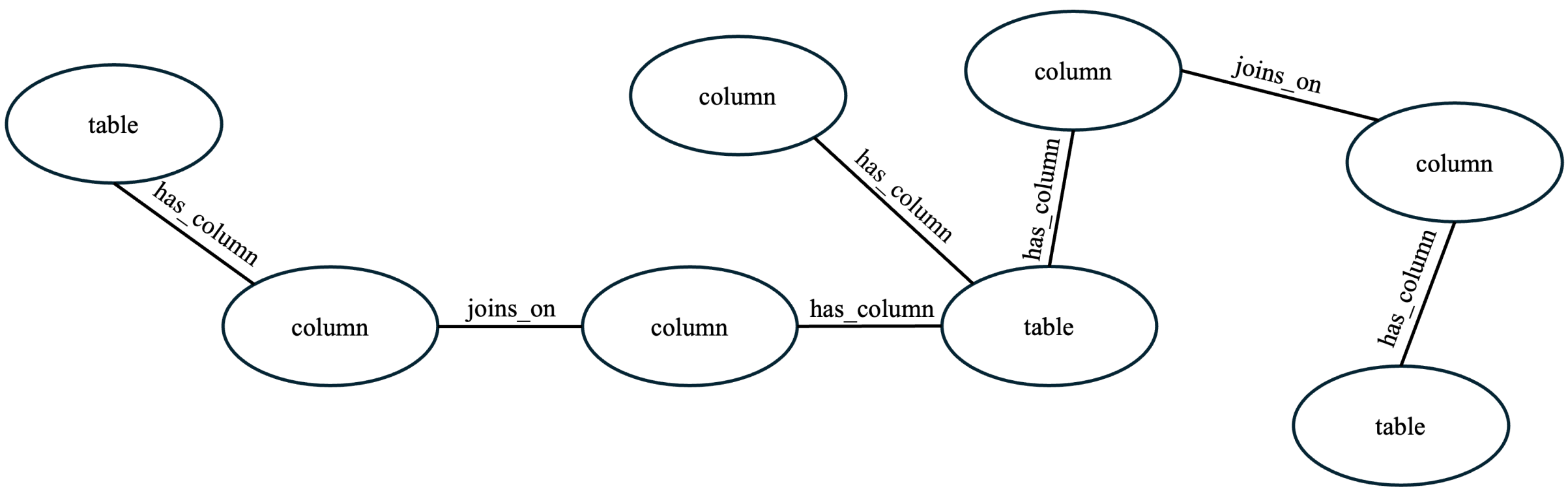

*Figure 4. Conceptualization of Knowledge Graph based on Database Schema.*

We store all KG nodes in a vector-store using the package ChromaDB 1.3.4 and the OpenAI embedding model text-embedding-3-large, representing table and column names along with their descriptions as embedding vectors. Each node is stored together with its metadata, including its identifier, type, name, and description in the vector-store. This configuration enables retrieval of relevant nodes based on user input, supporting semantic matching and reasoning over the KG. Using Chroma's *similarity_search* function, the system can identify relevant nodes based on semantic similarity rather than exact word matches. Nodes retrieved from the vector store serve as entry points into the knowledge graph, from which the system performs graph traversal to explore related tables and columns, ultimately identifying join paths and constructing a subgraph for the mediation process.

## 5.3 Demonstration

In this section, we highlight how KG-based mediation can effectively bridge the gap between user intent and complex enterprise data. We illustrate this through a structured three-step approach that interprets (imperfect) user prompts, maps them to relevant KG elements, and generates a refined prompt aligned with the underlying database. First, the user prompt is split into subsections that represent distinct pieces of information or intent. For each subsection, we retrieve the most relevant KG nodes based on semantic similarity, leveraging the embeddings stored in our vector-store. Second, for each identified table node, we retrieve all its columns, and for each identified column node, we retrieve its parent table, ensuring comprehensive coverage of schema elements related to the user's intent. Third, for each table found, we traverse the KG to identify both direct and indirect join relationships, adding any intermediate tables or columns that may be necessary to construct a complete join path.

To illustrate the difference between KG-based and schema-based mediation, we consider the user prompt: *"Which audit requests and findings still need to be completed, including company name and ID?"* For the example prompt, the query is first split into distinct segments. For each segment, relevant nodes are retrieved from the vector store and then traversed in the knowledge graph to identify related tables, columns, and join paths. Through this traversal, the KG-based mediation correctly identifies the *audit_single_issues* table for the audit requests that are not yet completed (*audit_single_issues.issue_done = FALSE*) and links these entries to the corresponding companies via *basis_info.unit_id*. The resulting query retrieves both the type of each open audit request (*audit_single_issues.issue_type*) and the related company name (*basis_info.unit_name*) and ID (*audit_single_issues.unit_id*).

The resulting mediated prompt is: *List all open audit requests and findings (audit_single_issues.audit_type) from the audit_single_issues table that are not yet completed (audit_single_issues.issue_done = FALSE). Also, include the company name (basis_info.unit_name) and company ID (audit_single_issues.unit_id) by joining audit_single_issues with basis_info on unit_id*. In total, the KG-based method successfully connected 21 nodes with 30 edges, effectively capturing the relationships needed to answer the user's question. In contrast, the schema-based approach, by comparison, identified the same tables (*basis_info*, *audit_status*, *audit_single_issues*) and 76 columns in total. However, it was unable to identify the correct tables and instead selected incorrect ones in response. By contrast, the resulting query returned only information from the *audit_status* table, which corresponds to the higher-level audit period rather than individual audit requests and their associated companies. This approach resulted in the wrong prompt: *Search the audit_status table for the value of P8A_AUDIT_NAME for all existing audits.* Figure 5 highlights the portion of the KG relevant to the query, illustrating the retrieved tables and columns along with their direct and indirect join relationships, and visually demonstrating how the KG-based mediation captures the connections from the source data.

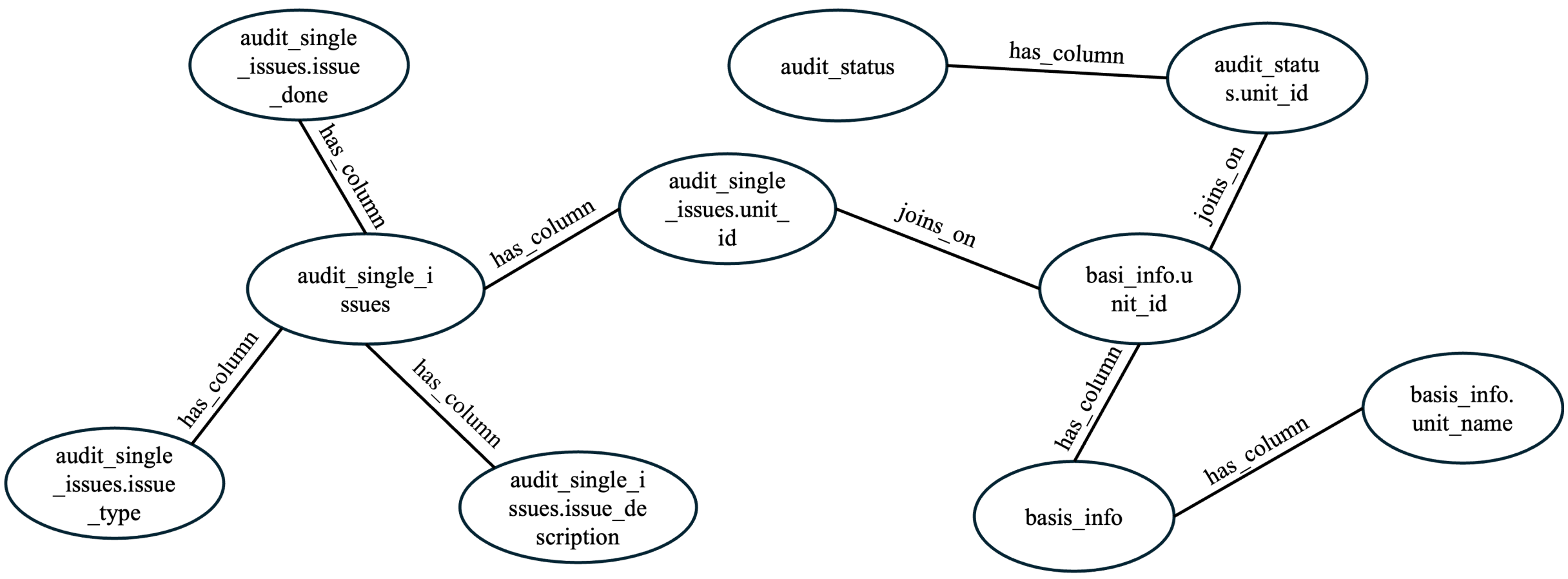

*Figure 5. Demonstration of Knowledge Graph.*

# 6 Evaluation

Our evaluation of the mediation unit is conducted using a comprehensive dataset based on log-data of user interactions with our enterprise data assistant from the first design cycle. This dataset encompasses data from 29 employees of the CFO unit of our industry partner, resulting in N=158 interactions. For our analysis, we use the semantic mediation unit to rewrite the real user queries from our dataset, with knowledge provided either through schema-based mediation or through KG-based mediation. Evaluation of the results is conducted using two complementary approaches.

First, we are using an LLM-as-a-Judge to measure *Completeness* between the two prompts, referring to whether all intents and key information from the original prompt remained. We apply completeness following Alabool (2025) as an LLM's ability to provide sufficient and relevant information in response to user inquiries, preserving all critical aspects to avoid compromising decision-making. The LLM was prompted to judge whether the two prompts preserve the same intent or lose details, outputting a *COMPLETED*/*NOT COMPLETED* label along with an explanation. A binary judgment is used because any missing information can mislead the system entirely, producing a false direction rather than a partially correct answer when querying the database. To mitigate potential model bias and improve consistency in judgments, we included three few-shot examples in the evaluation prompt that illustrate both completed and non-completed cases. We implemented this procedure using a Pydantic schema that enforces a structured output with the following fields: (1) the original user prompt, (2) the mediated prompt, (3) a Boolean variable indicating whether the mediated prompt should be labelled as completed, and (4) a detailed explanation specifying why the intent of the mediated prompt is considered completed or not completed with respect to the original prompt. Second, two data analysts from our industry partner evaluated in two iterations whether all necessary database information (tables, columns, filters) was correctly added to the prompt (*True Positives TP*), only some database information was added correctly to the prompt (*False Negatives FN*), or if database information added to the prompt was incorrect (*False Positives FP*). In the first iteration, one analyst conducted the review; in the second iteration, the other analyst reviewed these results. They then compared their assessments, discussed discrepancies, and resolved any conflicts before finalizing the annotations. Based on these, we calculated *Precision*, *Recall* and *F1-Score*. Table 1 summarizes the results of our evaluation using enterprise log-data (*N=158*) from users (*N=29*) in the CFO unit of our industry partner, based on interactions with our enterprise data assistant designed in the first design cycle. For our mediation unit, we evaluated two knowledge retrieval strategies: the baseline schema retrieval using an embedding model or our proposed KG retrieval mechanism. Prompt refinement is done using two LLMs: GPT-4.1 and Claude Sonnet 4. Completeness was determined using an LLM-as-a-Judge approach to classify if intents and key information were preserved. Based on the results of the two human annotators, *Precision*, *Recall* and *F1-Score* are calculated using the standard definitions in terms of true positives, false positives, and false negatives. For the baseline, we only report intent completeness from the LLM-as-a-Judge approach. Because this condition does not have access to schema or KG information and therefore does not explicitly introduce database elements, TP/FP/FN-based metrics are not applicable and therefore not included in Table 1.

| | **Baseline** | | **Semantic Mediation** | | | |
|---|---|---|---|---|---|---|
| **Model** | **GPT-4.1** | **Claude Sonnet 4** | **GPT-4.1** | | **Claude Sonnet 4** | |
| **Variant** | Baseline | | Schema | KG | Schema | KG |
| Word Count of Refined Prompts | M=18; SD=12 | M=17; SD=11 | M=29; SD=12 | M=30; SD=14 | M=37; SD=23 | M=33; SD=20 |
| Completeness | 91,14% | 85,33% | 89.24% | **93.04%** | 84.52% | **86.45%** |
| Precision | Not applicable | | 69.05% | **90.67%** | 75.17% | **87.23%** |
| Recall | Not applicable | | 73.11% | **88.31%** | 92.56% | **93.52%** |
| F1-Score | Not applicable | | 71.02% | **89.47%** | 82.17% | **90.29%** |

*Table 1. Results of the Simulation-based Evaluation.*

Across both LLMs (GPT-4.1, Claude Sonnet 4), the KG-based mediation consistently outperformed the schema-based mediation architecture and the baseline. Completeness was slightly higher for KG-based mediation, indicating that this approach provided a more compressed and relevant context window for the LLM. With GPT-4.1, the KG-based approach achieved a *Precision* of 90.67% (Schema: 69.05%), a *Recall* of 88.31% (Schema: 73.11%), and a *F1-Score* of 89.47% (Schema: 71.02%). Using Claude Sonnet 4, the KG-based mediation also showed a significant increase in *Precision* to 87.23% (Schema: 75.17%), while *Recall* remained comparable at 93.52% (Schema: 92.56%). The *F1-Score* for Claude Sonnet 4 improved to 90.29% (Schema: 82.17%). Annotator agreement was 98.73% (KG-based) and 98.10% (Schema-based) for GPT-4.1 and 95.57% (KG-based) and 96.84% (Schema-based) for Claude Sonnet 4, with discrepancies resolved during the second annotation iteration. In terms of processing time, schema-based mediation averaged 1.1 seconds, whereas KG-based mediation was slightly slower with 2.7 seconds. From the vector-store retrievals, the KG yielded an average of 54 nodes (SD=30) and 99 edges (SD=41) per query, while the corresponding database schemas included 91 columns on average (SD=19). We observed that KG-based mediation produced modestly higher word counts for some prompts because it more frequently surfaced explicit join specifications for GPT-4.1, which were typically correct and actionable. In contrast, Claude Sonnet 4 integrated joins more consistently, leading to similar word counts. Comparing the mediated conditions to the baseline regarding completeness reveals that both mediation approaches preserve user intent at least as well, even in the presence of additional database content. For GPT-4.1, KG-based mediation achieves a completeness of 93.04% compared to 91.14% in the no-mediation baseline, while also introducing explicit and largely correct database references. This shows that the model's performance remains high despite being exposed to richer database content. Schema-based mediation slightly reduces completeness (85.33%) but still maintains a high level of intent preservation. For Claude Sonnet 4, similar completeness across conditions indicates that introducing explicit schema elements does not compromise intent preservation. Notably, only semantic mediation can incorporate such schema information without inducing false information or hallucinations, demonstrating that semantic mediation reliably guides the model while maintaining user intent, even when additional database content is provided. It should be noted that in some cases prompt mediation resulted in only minimal changes to the original prompt due to a lack of domain and database knowledge, and this behaviour is not captured by the completeness metric.

In a preliminary controlled test using a random subset of 30 interactions for end-to-end trials with our enterprise data assistant, some prompts (true positives) proceeded without any additional schema discovery, reducing latency and accelerating result retrieval, while others still required explicit schema discovery to produce the correct results. In contrast, FN cases either required explicit schema lookup or returned correct but broader results that needed subsequent narrowing. This behaviour is reflected in the evaluation metrics: higher F1 scores for KG-based mediation correlate with higher end-to-end accuracy, as the mediator more consistently provides complete and precise information to the assistant. On complex database schemas, the schema-based mediation was more prone to erroneous or hallucinated joins; once such assumptions were encoded in the mediated prompt, downstream query planning and execution reliably failed. To address this, the next design cycle will leverage a complete, formal, curated dataset in collaboration with our industry partner. Given our insights and interaction log data, the dataset will be informed by imperfect user prompts rather than solely relying on idealized or artificially generated queries, ensuring that the evaluation reflects realistic user invocations. In summary, our observations align with the quantitative gains in *Precision*, *Recall*, and *F1-Score* for KG-based mediation and highlight the value of constraining join options at retrieval time. Reliable KG performance depended on identifying the right entities and relations, which was achieved via the vector-store that semantically matched user phrasing to canonical KG nodes. Because the database was thoroughly curated in the first design cycle, embedding-based indexing was straightforward for both the schema-based and KG-based methods. Finally, control checks with non-database prompts (TNs) confirmed that neither approach injected additional or incorrect knowledge.

# 7 Discussion

The widespread integration of enterprise data assistants has the potential to allow users at different skill levels to interact with data through natural language. However, despite the notable strengths of LLMs, including domain knowledge remains indispensable.

A promising approach to mitigate this gap is the provision of a semantic mediator that bridges the gap between user intent and source data characteristics. Unfortunately, knowledge provision can be limited, and excessive information can overwhelm LLMs, ultimately hindering their ability to mediate effectively. Drawing on research on mediators, our study addresses this gap by investigating how knowledge can be modelled, extracted, and used in LLM-based enterprise data assistants. We conducted a simulation-based evaluation which used enterprise prompts from users and designed a mediation layer that received knowledge either from database schemas or from a KG. Our results suggest that KG-based mediation (vs. schema-based mediation) reduces the likelihood of missing or incorrect information, better captures users' specific intents and the necessary relationships between knowledge pieces, and in turn more effectively supports data-driven decision with LLM-based data assistants. Additional analyses revealed that the KG helps to better represent database semantics and to incorporate more analytical instructions into prompts. Moreover, this mitigates the likelihood of including wrong information into prompts and helps users find the right information independent of their skill levels. To conclude our second design cycle, we propose our design principle of a KG-based semantic mediation layer for enterprise data assistants:

*When designing LLM-based enterprise data assistants (Aim), the artifact should incorporate a Knowledge Graph-based mediation layer that operates on and augments user prompts (Context). This unit acts as an intermediary, interpreting user intentions and representing them into context-aware source data representations (Mechanism). By embedding this mediation unit in the system, it reduces dependence on manual prompt engineering, enhances data retrieval efficiency, and bridges the gap between users' intentional goals and the underlying system representations (Rationale).*

Our results offer various theoretical and practical contributions, which we will discuss in the next sections.

## 7.1 Theoretical contribution

Our study makes three theoretical contributions to the literature which we will discuss below. First, it extends prior research on the role of mediators in IS by offering insights into how the delivery of knowledge, rather than the conceptualization of knowledge itself, can impact interaction with LLM-based enterprise data assistants. While existing research has predominantly focused on the curation of knowledge relationships and the technical procedures for effective knowledge retrieval (e.g., based on semantic similarities or specific graph retrieval languages) our work shifts the focus to knowledge delivery to reduce the user intent gap. Our findings show that domain-specific and data source relationships mitigate imperfect user prompts and clarification needs, highlighting the importance of knowledge inclusion in the design of enterprise data assistants. Second, we demonstrate that the provisioning of excessive context and unconnected knowledge undermines LLM performance and increases the potential for false assumption. While LLMs are often praised for their generalizability, our findings indicate that providing just the right context outperforms relying on large context windows. Third, our study provides insights into the interplay of vocabulary, vagueness and ambiguities when using an LLM-based enterprise data assistant. With the growing attention to imperfect prompts, the integration of knowledge excels assistants' capabilities beyond technical mechanisms such as clarification and multi-query techniques. Taken together, our findings indicate that ambiguities in language are not universal; instead, they depend on the intent of users, the domain the IS represents, and the capability to mediate between them.

## 7.2 Practical implications

Finally, we provide actionable insights and findings for developers and enterprises aiming to implement LLM-based enterprise data assistants, particularly in the context of data retrieval for a broader group of users, including those who may be unfamiliar with SQL or the underlying data structures. Since our results suggest that imperfections can reduce data assistant performance, organizations should prefer the explicit implementation of knowledge within their assistants, as it allows users to retrieve the information that they intend to access. Moreover, our findings underscore the importance of considering the role of users and the domain characteristics, beyond their experience with data analysis tools and LLMs. Developers and organizations should understand their target users and customize their assistants based on individual users and data characteristics.

## 7.3 Limitations and future research

Several limitations of our research highlight opportunities for future research. First, our study focuses on a single-enterprise setting, the CFO unit of our industry partner, and a finite set of 158 interactions from 29 users. While this scope enabled careful control of context and grounding in the real-world, it may limit the generalizability of our insights. Replication across industries, organizational functions, and larger prompt corpora would help assess boundary conditions and strengthen external validity. Second, our simulation-based evaluation depends on log-data, validated by human annotators instead of real-time user input. Future work could incorporate field deployments, controlled A/B studies, and qualitative research to triangulate findings, capture longitudinal effects, and assess changes when exposing users to a semantic mediator. This limitation will be addressed in the next design cycle through user-centred, end-to-end evaluations to validate the proposed artifact and to examine how user interactions and prompt formulations evolve in response to mediation. Third, the performance of the overall assistant still depends on the KG, its coverage, and maintenance processes. Although this reduces ongoing effort compared to extensive few-shot prompt engineering, production environments require systematic governance for continuous growth, versioning, and change management. Research should explore lifecycle practices, tool support, and organizational roles for long-term KG stewardship. Fourth, advancing the KG itself offers substantial potential. Extending the graph concept to include richer constraints, and explicit domain vocabulary could deepen the mediator's capacity. Finally, future research could refine KG-based mediation or extend it into a hybrid mediation that adapts based on user intent, task context, prompt complexity, or access rights. While we did not systematically vary network density, our results demonstrate that KG-based mediation is effective for the specific configuration used in this study. The knowledge graph is automatically generated and reflects the structure of the underlying database, which may differ across application domains. A promising avenue for future work is to investigate how graph retrieval strategies and reasoning perform under varying network structures and densities. Studying how different subgraph granularity and windowing strategies affect prompt mediation, latency, and costs would inform practical design trade-offs.

# 8 Conclusion

This study investigates the impact of two architectural variants and the provision of knowledge on user prompts in the context of LLM-based enterprise data assistants. Our results suggest that a semantic mediation layer, combined with embedding models and KGs, helps mitigate the imperfections of user prompts, which in turn positively influences the performance of the enterprise data assistant, highlighting the importance of modelling and tailoring IS to users. Thus, the study contributes to the literature on the design of LLM-based enterprise data assistants from a user perspective by emphasizing the role of mediators and knowledge. Taken together, these insights encourage the design of enterprise assistants that externalize domain specifics, meet users where they are, and continually evolve as enterprise knowledge grows.